\documentclass[aps,prl,twocolumn,groupedaddress,showpacs,amsmath,amssymb]{revtex4}

\newcommand\thintilde{{\lower.74ex\hbox{\mathtt{\char`\~}}}}
\usepackage{graphics}
\usepackage{color}
\usepackage{graphicx}% Include figure files
\usepackage{natbib}
\newcommand{\un}[1]{\ensuremath{\, \mathrm{#1}}}

\begin{document}

\title{Ultra-Fast Physical Generation of Random Numbers Using Hybrid Boolean Networks}

\author{David P. Rosin,$^{1,2}$ Damien Rontani$^1$ and Daniel J. Gauthier$^1$}

\affiliation{$^1$Department of Physics, Duke University, Department of Physics, Durham, North Carolina 27708, USA\\
$^2$Institut f\"ur Theoretische Physik, Technische Universit\"at Berlin - Hardenbergstr.~36, D-10623 Berlin, Germany}
\thanks{}
\date{\today}

\begin{abstract}
We describe a high-speed physical random number generator based on a hybrid Boolean network with autonomous and clocked logic gates, realized on a reconfigurable chip. The autonomous logic gates are arranged in a bidirectional ring topology and generate broadband chaos. The clocked logic gates receive input from the autonomous logic gates so that random numbers are generated physically that pass standard randomness tests without further post-processing. The large number of logic gates on reconfigurable chips allows for parallel generation of random numbers, as demonstrated by our implementation of 128 physical random number generators that achieve a real-time bit rate of $12.8\un{Gbit/s}$.
\end{abstract}

\pacs{05.45.Gg, 64.60.aq, 05.45.-a}

% 05.45.Gg : Application of Chaos.
% 64.60.aq : Networks in phase transitions
% 89.75.Da : Scaling Phenomena in complex systems
% 05.45.-a : nonlinear dynamics: replacing the pac number above
\maketitle

True random numbers guarantee the integrity of cryptographic protocols used in secure communications \cite{ASM07} and the accuracy of Monte-Carlo simulations used in climate or biomedical sciences \cite{MET49}. They can only be extracted from physical entropy sources, such as quantum fluctuations \cite{KWY10,ROY11} and electrical noise \cite{JUN99}, as opposed to pseudo-random numbers that are obtained from deterministic mathematical algorithms. Recently, it was found that chaotic dynamics in the presence of constant perturbation by microscopic noise can also be used for the physical generation of random numbers \cite{HAR12,KAN09}. With this approach, real-time rates of $\sim2\un{Gbit/s}$ have been achieved \cite{UCH08,UCH11_a}.

The generation of true random numbers presents challenges because of intrinsic bias (non-identical proportion of ``0'' and ``1'') and correlations originating from the physical source. Both effects diminish the quality of the random numbers, resulting in failure of statistical randomness tests \cite{NIST}. Bias and correlations can be reduced using post-processing, such as by combining multiple bit streams from identical uncoupled systems or hashing random bits from the bit stream produced by a single device \cite{UCH08,JUN99}. The downside of these approaches is an increase in the number of devices or a decrease in the bit rate, respectively. In most studies of ultra-fast random number generation, the post-processing is performed off-line \cite{KAN09} because of technological limitations so that the actual generation rate of high-quality random numbers is much lower. Other challenges include scalability to higher rates and interfacing with  computing and communication architectures, which have triggered recent developments of monolithically-integrated devices \cite{UCH11_a}.

Chaotic photonic devices have dominated the research on the high-speed physical generation of random numbers because of their large bandwidth, which enables very high data rates \cite{UCH08}. Recently, experiments have demonstrated that broadband chaos can also be generated by circuits of autonomous logic gates, which do not rely on a clock. These electric circuits are also promising candidates for random number generation because they can be implemented on compact electronic chips \cite{ZHA09,DIC07}, which are inexpensive and highly integrable in communication architectures. However, due to large bias in their output voltage, they fail statistical randomness tests. They are a physical implementation of a generic class of complex systems known as autonomous Boolean networks (ABNs) \cite{GLAS96,GLAS97}. ABNs consist of nodes with Boolean outputs that evolve continuously in time according to a Boolean function of the Boolean inputs. The electronic realization of ABNs leads to the generation of analog output values because of the continuous switching between the Boolean voltages.

Field-programmable gate arrays (FPGAs) are ideally suited to realize large-scale experiments on Boolean networks \cite{ROS12}. Typically, $10^5$ to $10^6$ logic gates can be assigned to execute any Boolean function and wired flexibly to constitute any desired network topology. Networks that generate random numbers can be implemented in parallel to multiply the bit rate, and inexpensive interfaces can be used for the high-speed connection to a computing or communication architecture.

In this Rapid Communication, we show that random numbers can be generated physically using a hybrid Boolean network (HBN), which consists of the combination of an ABN in the chaotic regime integrated with a single clocked node. First, we analyze the autonomous part of the HBN and describe its transition to chaos. Above a certain network size, we find that the autonomous part generates high entropy. Then, we discuss the clocked part of the HBN that mixes the dynamics of the autonomous part to reduce bias and correlations. As a result, our HBN generates high-quality physical random numbers that pass standard randomness tests successfully without further post-processing at a rate of $100\un{Mbit/s}$. Finally, by implementing 128 uncoupled HBNs in parallel on a single FPGA, we achieve a real-time rate of $12.8\un{Gbit/s}$.

\begin{figure}[t]
\begin{center}
 \resizebox{7.8cm}{!}{\includegraphics{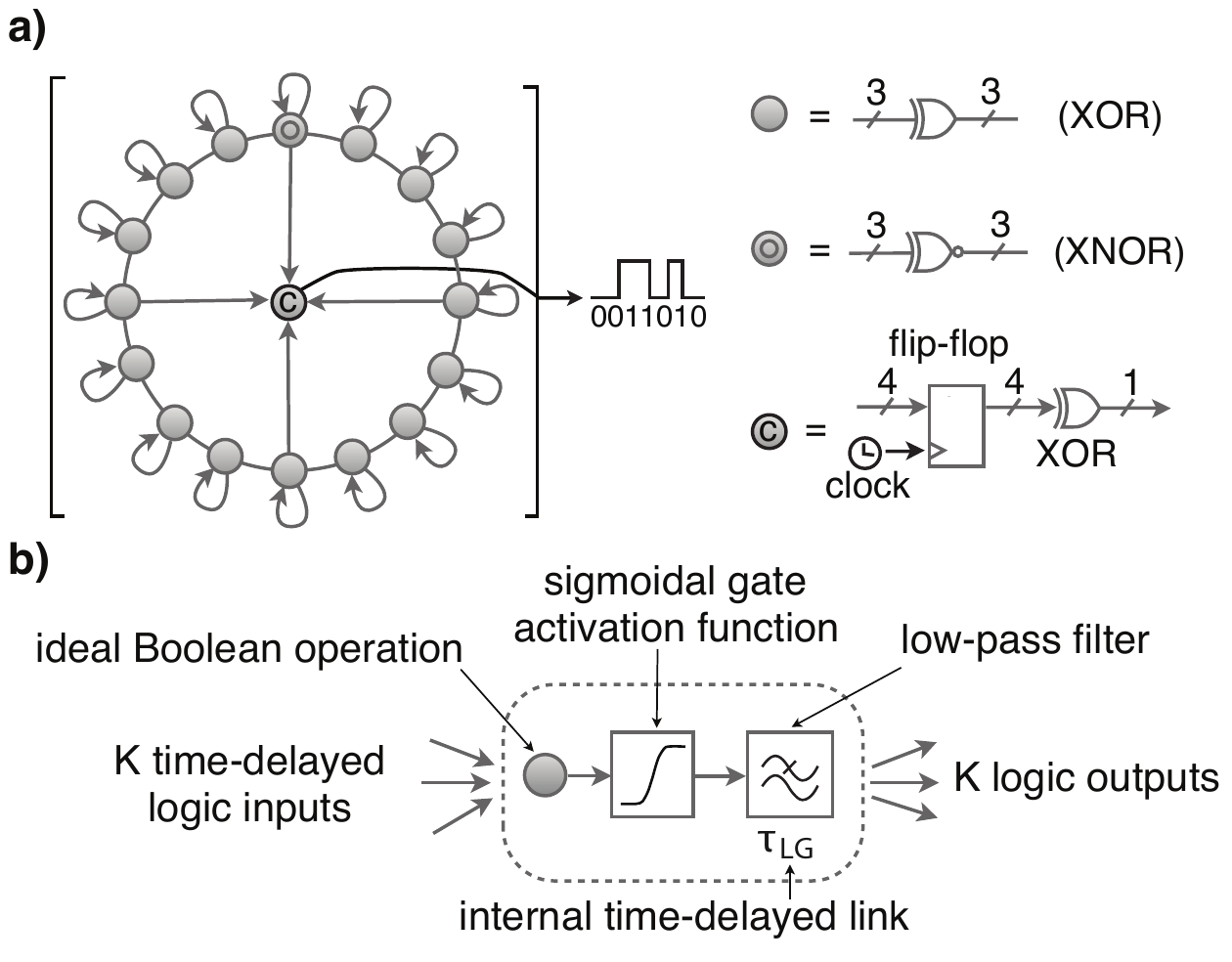}}
\end{center}
\caption{\label{fig:Figure1} a) Hybrid Boolean network (HBN) composed of $N=16$ autonomous nodes and one clocked node. Nodes execute XOR and XNOR logic functions with three (four) inputs and three (one) outputs for the autonomous (clocked) node. The clocking of the central node is realized using four clocked flip-flops. Links without arrows are bidirectional. The output of the HBN is a random bit stream. b) Non-ideal behaviors present in an electronic logic gate.}
\end{figure}

Our HBN is illustrated in Fig.~\ref{fig:Figure1}a; its autonomous part is an ABN with a topology previously proposed in Ref.~\cite{BAE08}. The ABN consists of $N$ nodes with three inputs and outputs assembled in a bidirectional ring topology with nearest-neighbor coupling and self-feedback, where $N-1$ nodes execute the exclusive-OR (XOR) Boolean function and one node executes the Boolean inverse of the XOR (the XNOR Boolean function).

When operated autonomously, logic gates are subjected to various non-ideal effects, as illustrated in Fig.~\ref{fig:Figure1}b. Specifically, their dynamics are affected by low-pass filtering, a finite gate propagation time $\tau_\mathrm{LG}$ ($\tau_\mathrm{LG}=280\pm10\un{ps}$ for the Altera Cyclone IV FPGA), and a continuous input-output relationship described by a sigmoidal gate activation function. The combination of these effects has significant impact on the dynamics of the ABN.

%%%%%%%%%%%%%%%%%%%%%%%%%%%%%%%%%%%%%%%%%%%%%%%%%%%%%%%%%%%%%%%%%%%%
%% Figure 1 - Random Boolean Network & Physical Model
%%%%%%%%%%%%%%%%%%%%%%%%%%%%%%%%%%%%%%%%%%%%%%%%%%%%%%%%%%%%%%%%%%%

The dynamical behavior of the ABN for a small number of nodes $N<5$ is a steady state, which involves node output voltages that are at intermediate values between low and high Boolean voltages. Consequently, the output voltages satisfy a self-consistency condition of the sigmoidal gate activation function of the logic gates.

In this steady-state regime, the dynamics are prevented from displaying oscillations because the propagation time around the network $N\tau_\mathrm{LG}$ is short. Here, oscillations with the lowest possible frequency $f=1/(2N \tau_\mathrm{LG})$ (corresponding to a single Boolean transition traveling around the ring network) are attenuated by the low-pass frequency filtering characteristics of the nodes. That is, the frequency-dependent gain is less than one for that frequency and all higher modes. For the ABN to bifurcate to non-steady dynamics, the propagation time must be larger, which is achieved by increasing $N$.

%%%%%%%%%%%%%%%%%%%%%%%%%%%%%%%%%%%%%%%%%%%%%%%%%%%%%%%%%%%%%%%%%%%%
%% Figure 3 - First Instability & jitter
\begin{figure}[b]
\begin{center}
 \resizebox{7.75cm}{!}{\includegraphics{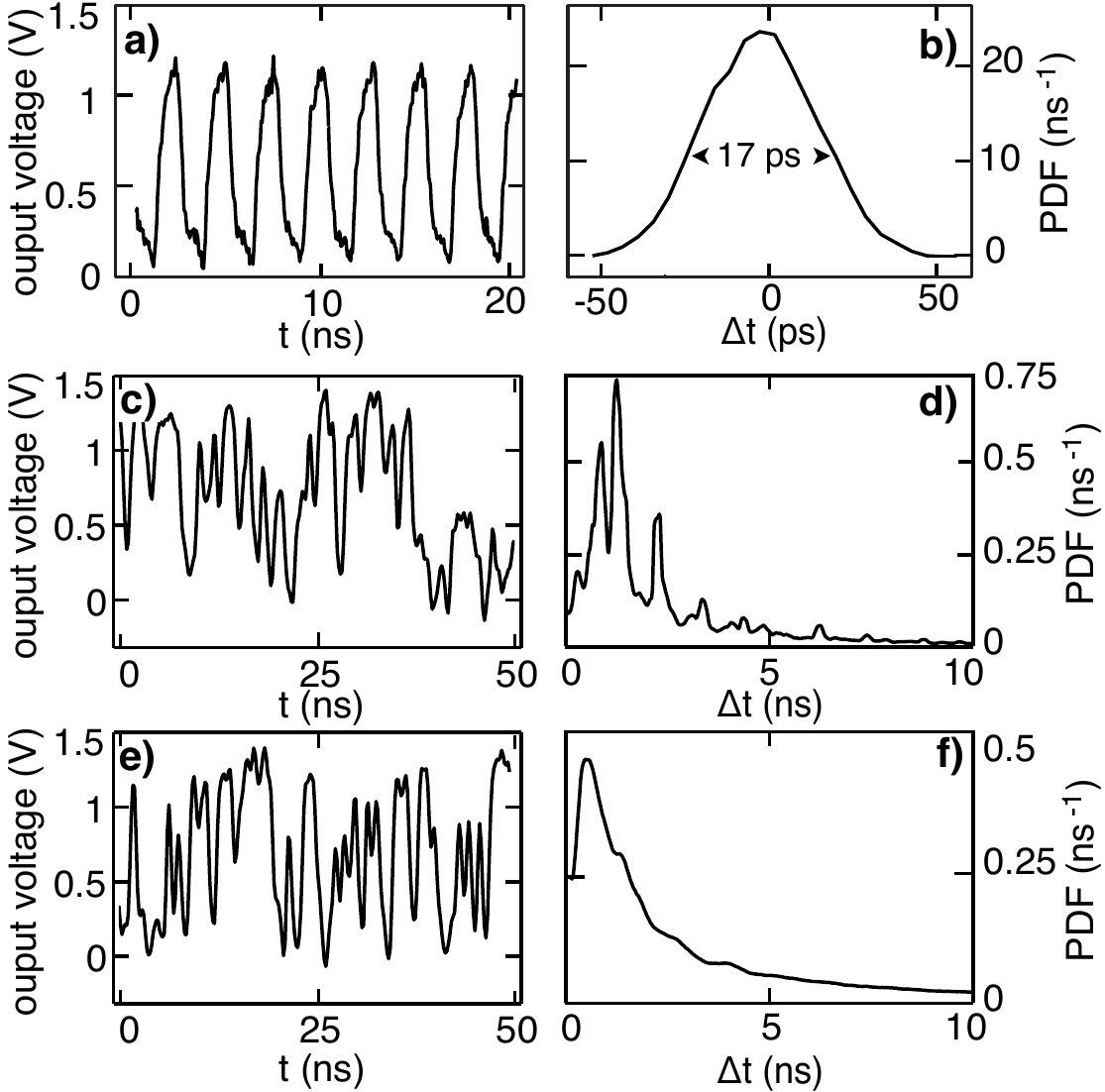}}
\end{center}
\caption{\label{fig:Figure2}
a) Temporal evolution of the network for $N=5$ measured from one autonomous node, displaying oscillatory dynamics. b)  Distribution of the time differences $\Delta t$ between two consecutive Boolean transitions for $N=5$, quantified by the probability density function (PDF) centered at zero. c) Temporal evolution and d) PDF for $\Delta t$ for $N=6$ when the ABN displays chaos. e) Temporal evolution and f) PDF for $\Delta t$ for $N=16$ when the ABN also displays chaos. The Boolean networks are realized on the FPGA Altera Cyclone IV EP4CE115. Data and Boolean transitions are sampled at 5 GSa/s with 12-bit quantization and 8 GHz analog bandwidth using oscilloscope DSO 90804A.}
\end{figure}
%%%%%%%%%%%%%%%%%%%%%%%%%%%%%%%%%%%%%%%%%%%%%%%%%%%%%%%%%%%%%%%%%%%

For $N=5$, the ABN displays periodic oscillations with frequency $f\approx1/(2N \tau_\mathrm{LG})$, as shown in Fig.~\ref{fig:Figure2}a. The oscillatory dynamics originate from the low-pass filter characteristics of the logic gates. The first oscillation mode with frequency $f=1/(2N \tau_\mathrm{LG})$ is amplified and the next mode with frequency $f=1/(N \tau_\mathrm{LG})$ and all higher modes are still attenuated. Hence, the ABN has a size corresponding to a propagation delay that allows for exactly one Boolean transition to propagate.

%inverter oscillators for random number generation
As shown in Fig.~\ref{fig:Figure2}b, the periodic oscillations generated by the ABN for $N=5$ have timing jitter, which is a random fluctuation of the phase due to thermal, shot, and flicker noise from each logic gate \cite{LEE98}. The jitter in the oscillatory dynamical regime of ABNs can be exploited for physical random number generation, as detailed by previous studies on inverter oscillators \cite{SUN07,WOLD09}. In one approach, random numbers are generated using multiple unidirectional ring oscillators built from $\sim1300$ autonomous inverter gates together with multiple clocked XOR logic gates. However, this standard approach has been proven to be flawed \cite{DIC07}.

%reason for chaos
For $N>5$, the ABN displays chaotic dynamics corresponding to irregular fluctuation between the two Boolean voltage levels \cite{ZHA09} as shown in Fig.~\ref{fig:Figure2}c and e for $N=6$ and $N=16$, respectively. The frequency spectrum of the chaos (not shown) extends to $\sim600\un{MHz}$ at the $10\un{dB}$ drop-off point.

Chaos in the ABN is a result of our choice of logic functions, high connectivity, heterogeneity in propagation delays, and the autonomous mode-of-operation. The three-input XOR and XNOR logic functions have maximum Boolean sensitivity, meaning that a Boolean transition in any of the three inputs induces a Boolean transition in the output \cite{KAU04}. Furthermore, whenever a node creates a Boolean transition, it is distributed to its two neighbors and to itself. The heterogeneity in the propagation delays implies that transitions occur with a relative time shift, as opposed to occurring simultaneously as in synchronous Boolean networks \cite{POM09}. Heterogeneities are caused by the inherent variability and state and history dependence of the propagation delay of logic gates \cite{CAV10,SOC07} and path differences between the on-chip links. Consequently, a transition in the output of one autonomous node can cause three subsequent transitions, themselves inducing nine transitions. This cascading process leads to an accumulation of transitions with a generation rate that is bounded by the cut-off frequency of the low-pass filter of the logic gates.

The chaos can be characterized by the distribution of time intervals between two consecutive Boolean transitions, as shown in Fig.~\ref{fig:Figure2}d and f. These distributions are $\approx200$ times broader than the jitter distribution in the oscillatory regime (Fig.~\ref{fig:Figure2}b). Consequently, more randomness can be extracted from the dynamics in the chaotic regime than from jitter in the periodic regime shown in Fig.~\ref{fig:Figure2}a. However, the distribution of transitions for $N=6$ presents significant fluctuations (Fig.~\ref{fig:Figure2}d) that can introduce undesired statistical properties of the random numbers, when compared to those provided by a network with $N=16$ nodes (Fig.~\ref{fig:Figure2}f).

We characterize the irregularity of the chaotic waveform generated by an autonomous node in the HBN using the Shannon entropy \cite{Cover91}. The quality of the random numbers produced by the clocked node in our HBN is, however, assessed with standard randomness tests. To compute the entropy, we sample periodically the voltage of an autonomous node with period $T_s = 10\un{ns}$ and 1-bit quantize to obtain a bit stream representative of the dynamics of the autonomous part of our HBN.

\begin{figure}[t]
\begin{center}
 \resizebox{7.75cm}{!}{\includegraphics{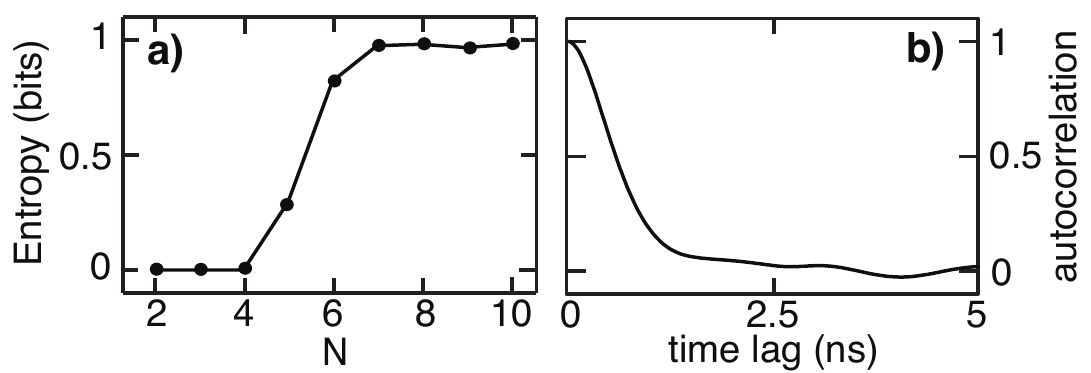}}
\end{center}
\caption{\label{fig:Figure3}
a) Entropy of the analog waveform generated by an autonomous node in the HBN as a function of $N$ defined by $H(X) = -\sum_{x_n=\{0,1\}} p_{X}(x_n)\log_2 p_{X}(x_n)$ , where $X$ is the 1-bit-quantized voltage sampled at $10\un{ns}$ intervals and $p_{X}(x_n)$ is the associated probability density of $X$. b) Autocorrelation of the chaotic time series of one autonomous node for $N=16$.}
\end{figure}

The entropy measured from one autonomous node is shown in Fig.~\ref{fig:Figure3}a as a function of $N$. For $N<5$, the entropy is $H=0\un{bit/sample}$; for oscillatory dynamics with $N=5$, it is $H\approx (0.30\pm0.01)\un{bit/sample}$. In the chaotic regime with $N=6$, the entropy increases to $H\approx(0.82\pm0.01)\un{bit/sample}$, and for $N>7$ it is $H\approx(0.96\pm0.01)\un{bit/sample}$. The overall increase in entropy describes a dynamical transition with $N$ and is a consequence of the three different dynamical states. First, the steady state results in a constant bit stream that has zero entropy. Second, jitter in periodic dynamics creates a small uncertainty leading to a low level of entropy for $T_s = 10\un{ns}$. Third, the chaotic dynamics has entropy close to the maximum achievable value of $H=1\un{bit/sample}$ \cite{Cover91} when sampled with $T_s$.

%%%%%%%%%%%%%%%%%%%%%%%%%%%%%%%%%%%%%%%%%%%%%%%%%%%%%%%%%%%%%%%%%%%%
%% Figure 3 - Random Number Generation
%%%%%%%%%%%%%%%%%%%%%%%%%%%%%%%%%%%%%%%%%%%%%%%%%%%%%%%%%%%%%%%%%%%

To reduce the residual bias, we form an HBN that integrates the ABN with $N=16$ and a clocked node with four inputs (a standard number of inputs for logic gates on FPGAs). The ABN generates chaotic dynamics with a saturated entropy rate. The clocked node of the HBN realizes a 1-bit quantization operation. Such a quantization level is adequate for extracting randomness from the ABN Boolean chaos whose complexity originates from the timing of Boolean transition rather than the voltage amplitude \cite{ZHA09}. The clocked node also performs an XOR operation, which is known to reduce bias if the incoming bit streams are sufficiently uncorrelated \cite{UCH08}.  We find that the cross-correlation between every four nodes in the ABN leads to a normalized cross-correlation level below $0.075$. This is low enough for the clocked node to generate a single bit stream with sufficiently low bias to pass standard randomness tests. The clocked node includes a flip-flop that samples the autonomous nodes with a clock frequency of $100\un{MHz}$, as shown in Fig.~\ref{fig:Figure1}a. This sampling rate corresponds to the maximum transfer rate to the memory elements on the FPGA and results in our HBN generating random numbers at a bit rate of $100\un{Mbit/s}$. Higher bit rates are allowed by our network because of the short correlation time ($\approx 590\un{ps}$) of the chaotic dynamics generated by an autonomous node, as shown in Fig.~\ref{fig:Figure3}b. However, the memory transfer speed of our inexpensive FPGA platform precludes us from extracting this accessible entropy rate.

We implement the HBN-based random number generator on various FPGAs, such as the Altera Cyclone IV, Altera Stratix IV, Xilinx Virtex VI, and the CPLD Altera MAX II. For all of the different FPGAs, we assess the quality of the random numbers with $1\un{Gbit}$ of data using the NIST  and Diehard test suites \cite{NIST,DieHard} and observe consistently successful passes at a rate of $100 \un{Mbit/s}$. This shows that our approach is robust to changes in technology. Furthermore, the HBN-based random number generator requires only 17 logic gates, which is significantly less than the number needed for random number generators based on unidirectional ring oscillators ($>1{,}300$ \cite{SUN07}).

\begin{table}
\caption{\label{tab:table1}
Results of the 800-22 NIST test suite \cite{NIST} using $1\un{Gb}$ of data (1000 sequences of 1 Mb) generated by the 128 HBN-based random number generators implemented in parallel on an Altera Cyclone IV EP4CE115 FPGA. All tests are passed successfully because the P-value is larger than 10$^{-4}$ and the proportion is greater than 0.980.}

\begin{ruledtabular}
\begin{tabular}{cccc}
Statiscal Tests&P-value&Proportion&Result\\
\hline
Frequency                & 0.0856 & 0.991 & Success\\
Block frequency          & 0.7887 & 0.993 & Success\\
Cumulative sums          & 0.3191 & 0.988 & Success\\
Runs                     & 0.2954 & 0.989 & Success\\
Long runs                & 0.0081 & 0.992 & Success\\
Ranks                    & 0.1147 & 0.995 & Success\\
FFT                      & 0.4750 & 0.991 & Success\\
Nonoverlapping templates & 0.1445 & 0.983 & Success\\
Overlapping templates    & 0.6621 & 0.987 & Success\\
Universal                & 0.0288 & 0.990 & Success\\
Approximate entropy      & 0.5728 & 0.989 & Success\\
Random excursion         & 0.3694 & 0.982 & Success\\
Random excursion var     & 0.3917 & 0.982 & Success\\
Serial                   & 0.5544 & 0.987 & Success\\
Linear Complexity        & 0.4944 & 0.992 & Success\\
\end{tabular}
\end{ruledtabular}
\end{table}

The resulting random numbers generated by the clocked node of the HBN are non-deterministic because of the mixing property of chaos, where the electrical noise that induces jitter is mixed into the chaotic dynamics of the autonomous nodes. As a result, the chaos amplifies the entropy of the noise and, therefore, constitutes a larger entropy source than the noise itself \cite{HAR12,UCH11_a}.

The achievable random bit rate of a single HBN remains one order-of-magnitude below that of photonic systems. However, the small fraction of required logic gates is less than $0.02\%$ of the logic gate on an Altera Cyclone~IV FPGA, allowing us to parallelize thousands of random number generators, thus increasing the overall bit rate. Limited by the memory capacity of our FPGA, we implement 128 uncoupled HBNs in parallel that have independent temporal evolution and, together, generate 128 random bits per clock cycle. As we keep the sampling rate at $100\un{MHz}$, we achieve a cumulative bit rate of $12.8\un{Gbit/s}$. The cumulatively generated random numbers pass successfully the NIST tests using $1\un{Gbit}$ of data with results shown in Table~\ref{tab:table1}. The random numbers also successfully pass the Diehard tests; results are not shown here.

In addition to using the statistical test suites, we compute the $3\sigma$ confidence intervals for the bias $b$ and serial-correlation coefficient $\rho$ under the statistical null hypothesis $H_0$ of a perfectly uncorrelated, unbiased random number generator using $n=3\times10^9$ bits. They are given respectively by $\hat{b}\pm3\hat{\sigma}/\sqrt{n}=[-2.826,2.651]\times10^{-5}$ and $\hat{\rho}\pm 3\sqrt{(1-\hat{\rho}^2)/(n+1)}=[-4.177,6.777]\times 10^{-5}$ \cite{White57} with $\hat b$, $\hat\sigma$, and $\hat\rho$ being the test statistics for the bias, standard deviation, and serial-correlation coefficient, respectively. Because these two intervals contain $b=0$ and $\rho=0$, $H_0$ is not rejected with a $99.97\%$ confidence.

In conclusion, we generate physically high-quality random numbers with a hybrid Boolean network (HBN), which consists of an autonomous part that displays chaotic dynamics and a clocked part that mixes the chaotic dynamics. The HBN can be realized with a small number of logic gates on an FPGA, allowing for parallel implementation of many HBNs. The random numbers pass statistical randomness tests, as demonstrated for a single HBN and for 128 HBNs in parallel allowing for bit rates of $12.8\un{Gbit/s}$ without further processing. With current chip technology, we conjecture that tens of thousands of HBNs can be implemented in parallel because our implementation of 128 random number generators exploits less than $1\%$ of the available logic gates on typical FPGAs. Our approach using HBNs potentially opens the path towards Tbit/s physical random number generation on a single electronic device.

%%%%%%%%%%%%%%%

The authors gratefully acknowledge discussions of this work with Joshua Socolar, Hugo Cavalcante, Seth Cohen, and Kristine Callan. D.P.R., D.R., and D.J.G. gratefully acknowledge financial support of the U.S. Office of Naval Research Grant No. N00014-07-1-0734, the U.S. Army Research Office Grants No. W911NF-11-1-0451 and W911NF-12-1-0099. D.P.R. acknowledges the DFG for financial support in the framework of SFB910.

\bibliographystyle{apsrev4-1}
%\bibliography{references_PRErapid}
\end{document}